\documentclass[11pt, letterpaper]{article}
\usepackage[utf8]{inputenc}
\usepackage[margin=1.5cm]{geometry}
\usepackage{titlesec}
\usepackage{tabu}
\usepackage{enumitem}
\usepackage{amssymb}
\usepackage{xcolor}
\newlist{selectlist}{itemize}{2}
\setlist[selectlist]{label=$\square$,leftmargin=*,noitemsep,topsep=0pt}

\usepackage[style=numeric,citestyle=numeric-comp,sorting=none,backend=biber]{biblatex}
\usepackage{lmodern}

\usepackage{hyperref}
\hypersetup{
    colorlinks=true,
    linkcolor=blue,
    filecolor=magenta,      
    urlcolor=blue,
}

\titleformat{\section}[block]{\hspace{1em}\bfseries}{\thesection.}{0.5em}{} 
\titleformat{\subsection}[block]{\hspace{1em}}{\thesubsection}{0.5em}{}

\usepackage{graphicx}
\usepackage{caption}
\usepackage{cleveref}
\usepackage{upgreek}
\let\mu\upmu 

\Crefformat{figure}{#2Fig.~#1#3}
\Crefmultiformat{figure}{Figs.~#2#1#3}{ and~#2#1#3}{, #2#1#3}{ and~#2#1#3}
\Crefformat{equation}{#2Eq.~#1#3}
\Crefmultiformat{equation}{Eqs.~#2#1#3}{ and~#2#1#3}{, #2#1#3}{ and~#2#1#3}
\newcommand{\mendeley}{\href{http://dx.doi.org/10.17632/fy86gxm4mp.1}{repository link}}
\newcommand{\tu}[1]{\textup{#1}}

\begin{document}

\begin{flushleft}

\setlength{\parindent}{0pt}
\setlength{\parskip}{10pt}

\textbf{Article title}\\ 
Low-cost fluorescence microscope with microfluidic device fabrication for optofluidic applications

\textbf{Authors}\\ 
Nagaraj Nagalingam$^{a,*}$, Aswin Raghunathan$^{a,*}$, Vikram Korede$^{a}$, Edwin F. J. Overmars$^{a}$, Shih-Te Hung$^{b}$, Remco Hartkamp$^{a}$, Johan T. Padding$^{a}$, Carlas S. Smith$^{b}$ and Huseyin Burak Eral$^{a}$

$^*$denotes equal contribution

\textbf{Affiliations}\\ 
$^{a}$Process \& Energy Department, Delft University of Technology, Leeghwaterstraat 39, 2628 CB Delft, The Netherlands.\\
$^{b}$Delft Center for Systems and Control, Delft University of Technology, Mekelweg 2, 2628 CD Delft, The Netherlands

\textbf{Corresponding author’s email address and Twitter handle}\\ 
n.nagalingam@tudelft.nl (Nagaraj Nagalingam)

\textbf{Abstract}\\

Optofluidic devices have revolutionized the manipulation and transportation of fluid at smaller length scales ranging from micrometers to millimeters. We describe a dedicated optical setup for studying laser-induced cavitation inside a microchannel. In a typical experiment, we use a tightly focused laser beam to locally evaporate the solution laced with a dye resulting in the formation of a microbubble. The evolving bubble interface is tracked using high-speed microscopy and digital image analysis. Furthermore, we extend this system to analyze fluid flow through fluorescence$-$Particle Image Velocimetry (PIV) technique with minimal adaptations. In addition, we demonstrate the protocols for the in-house fabrication of a microchannel tailored to function as a sample holder in this optical setup. In essence, we present a complete guide for constructing a fluorescence microscope from scratch using standard optical components with flexibility in the design and at a lower cost compared to its commercial analogues.


\textbf{Specifications table}\\
\begin{tabu} to \linewidth {|X|X[3,l]|}
\hline  \textbf{Hardware name} & \textit{Low-cost Microscope - for laser assisted studies with in-house microfluidic device fabrication}
  \\
  \hline \textbf{Subject area} & %
  \begin{itemize}[noitemsep, topsep=0pt]
  \item \textit{Engineering and material science}
  \item \textit{Chemistry and biochemistry}
  \item \textit{Flexible and modular open source microscope}
  \item \textit{Educational method for microfluidic device fabrication}
  \end{itemize}
  \\
  \hline \textbf{Hardware type} &
  \begin{itemize}[noitemsep, topsep=0pt]
  \item \textit{Mechanical engineering and materials science}
  \item \textit{Protocol to chemical sample handling and preparation}
  \item \textit{Imaging tools}
  \item \textit{Laser assisted microscopy}
  \end{itemize}
  \\ 
\hline \textbf{Closest commercial analog} &
  Nikon's ECLIPSE Ti inverted microscope ($\approx \tu{\texteuro}\, 100,000-150,000$)
  \\
\hline \textbf{Open source license} &
  CC BY 4.0
  \\
\hline \textbf{Cost of hardware} &
  \texteuro\,46,429.05 (excluding laser and high-speed camera but including additional accessories such as optical breadboard, flow control device and 3D printer).\newline
  Refer Bill of Materials for the split-up of the costs.
  \\
\hline \textbf{Source file repository} & 
  Mendeley data: \href{http://dx.doi.org/10.17632/fy86gxm4mp.1}{http://dx.doi.org/10.17632/fy86gxm4mp.1}
  \\
\hline
\end{tabu}
\end{flushleft}

\newpage
\section{Hardware in context}

The evolution of optical microscopy has been sparked by the recent advancements in digital image processing and machine vision that have led to improved mobility and flexibility with microscopes through the use of open hardware and software \cite{10.1002/jemt.24200}. A fluorescence microscope is a non-invasive, visualization and optical measurement tool widely employed in the field of life sciences over the past several decades \cite{10.1038/nmeth817}. In recent times, a microscope combined with an excitation laser has gained traction in the optofluidics domain with an increasing number of applications that include cell sorting \cite{10.1186/1477-3155-2-5}, micro-robotics \cite{10.1089/soro.2019.0169}, microparticle synthesis \cite{10.1073/pnas.2005068117,10.1002/smll.201701804}, droplet drying \cite{10.1021/la8017858}, optical micromanupulation \cite{10.1063/1.2713164}, fluid microfilms and nanofilms \cite{10.1007/s00348-012-1279-3} and other microfluidic systems \cite{10.1039/C0LC00520G,10.1117/12.861757}. Commercially available microscopes often entail huge costs ($\tu{\texteuro}\, 100,000-150,000$ depending on the options and manufacturer) and offer limited flexibility with the design. With rapid technological advancements, microscopes have come with sophistications that require technical expertise for operation and maintenance \cite{10.1371/journal.pone.0167863}. Moreover, optofluidic applications generally require frequent modifications in the optical setup which makes the use of conventional microscopes difficult.

\bigskip

\noindent Several microscope designs have been developed specifically to perform laser-induced cavitation studies in microfluidic systems \cite{10.1007/978-981-287-278-4_6}. Rau et al. studied the hydrodynamic effects of cell lysis using a pulsed Nd:YAG laser beam \cite{10.1529/biophysj.105.079921}. In this setup, the source beam is bisected using a beamsplitter to both lyse the cell and illuminate the sample. Zwaan et al. performed controlled cavitation experiments by focusing an expanding Nd:YAG laser beam with a 40x objective to study the resulting planar bubble dynamics within different lab-on-a-chip device geometries fabricated using Polydimethylsiloxane (PDMS) \cite{10.1103/PhysRevLett.98.254501}. Similar experiments were carried out by Quinto-Su et al. and Sun et al. by focusing the Nd:YAG laser beam in microfluidic gaps of variable height created with coverslips and spacers \cite{10.1103/PhysRevE.80.047301} and a microtube with two different diameters \cite{10.1017/S0022112009007381}, respectively. Cavitation-based micropump was devised with a pulsed laser for its application in valve-less pumping \cite{10.1039/B806912C}. For the micropump, in addition to recording the bubble evolution with the high-speed camera, the effect of PDMS compliance was investigated together with the visualization of flow fields using the conventional PIV technique. Besides single component systems (e.g., aqueous dye), laser-induced cavitation of supersaturated aqueous solutions inside microwells have revealed crystal nucleation surrounding an evolving bubble \cite{10.48550/arxiv.2302.01218}. Techniques such as fluorescence-based thermometry \cite{10.1038/srep05445}, high-speed micro-PIV \cite{10.1088/0957-0233/15/10/003} and microparticle tracking velocimetry \cite{10.1063/1.2337506} when combined with cavitation experiments have proven to help with the characterization of the phenomena with the measurement of the associated parameters such as temperature and flow fields. Although many distinct microscopy configurations have been reported in the literature to perform laser-induced cavitation and PIV experiments, very limited emphasis is given to the design details, assembly of the experimental setup and the fabrication protocols for the microfluidic devices used.

\bigskip

\noindent In the present work, we delineate the construction of a low-cost customizable microscope using optomechanical components in conjunction with a pulsed laser to primarily study laser-induced microbubble dynamics within an in-house fabricated microchannel. In order to emphasize the flexibility of the devised setup, we conduct an experiment using the fluorescence-PIV technique achieved with a minor modification in the optics assembly. Furthermore, we demonstrate the fabrication procedure of PDMS microchannels used in the experiments. We limit our hardware design to the basic models of commercially available counterparts. However, with dedicated customization, its applications can be extended beyond the addressed laser-induced cavitation and PIV analyses.

\section{Hardware description}

\begin{figure}[t]
    \centering
    \includegraphics[width = 0.9\textwidth]{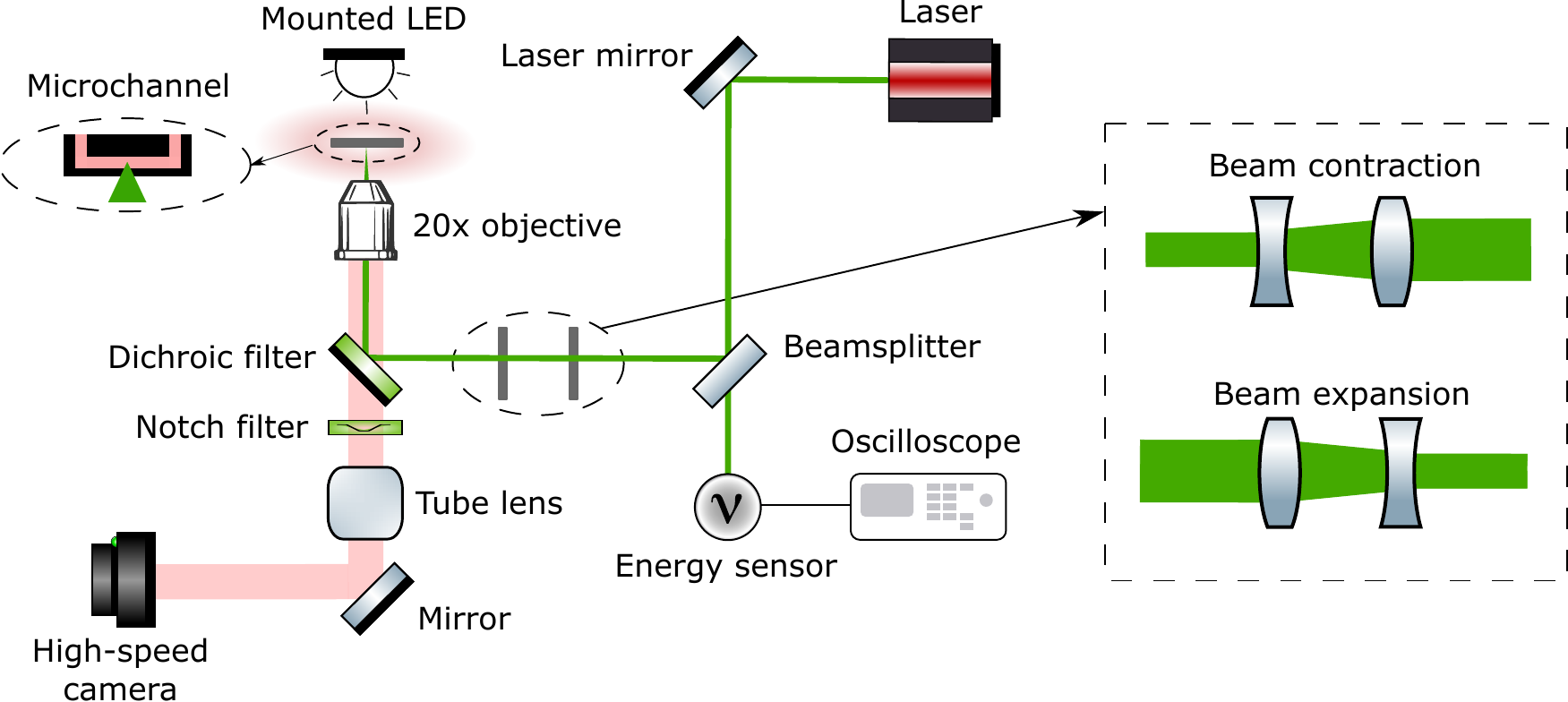}
    \caption{Architecture of the experimental setup. The sketch was made using a free vector graphics library for optics (\href{http://www.gwoptics.org/ComponentLibrary/}{\nolinkurl{http://www.gwoptics.org/ComponentLibrary/}}).}
    \captionsetup{justification=centering}
    \label{fig: Expt setup}
\end{figure}

A schematic of the optical architecture is provided in \Cref{fig: Expt setup}. Micro-sized vapor bubble is generated by heating the solution locally using a 532 nm, frequency-doubled, pulsed Nd:YAG laser having a beam diameter of 8 mm and a pulse duration of 4 ns. The beam path is guided from the laser source using a fully reflecting mirror and a partially reflecting beamsplitter. An energy meter connected to an oscilloscope is used to measure the pulse energy transmitted by the beamsplitter. The reflected laser pulse from the dichroic filter is directed towards a 3-axis translation stage where the microfluidic device is positioned. Depending on the desired laser beam diameter, a Galilean telescope arrangement (a set of convex and concave lenses separated by the sum of their focal lengths) can be constructed to either increase or decrease the laser beam diameter as indicated in \Cref{fig: Expt setup}. Alternatively, a beam expander with the required magnification range can be used for this purpose but is relatively more expensive. The flexibility of the design allows the placement of the telescope at any position between the source laser and the dichroic filter.

\bigskip

\noindent An inverted microscope arrangement consisting of an infinity-corrected 20x objective with a numerical aperture (NA) of 0.5 is used to focus the laser and form a microbubble. The energy intensity of the laser at the focal spot is estimated to be in the order of GW/cm$^{2}$. In this case, the same objective is also used to image the sample. A red LED light of 625 nm wavelength is used to illuminate the sample. In contrast to conventionally used white light, red light offers some advantages. Firstly, the entire light can pass through the dichroic filter, thereby providing better illumination at a lower intensity. Furthermore, the spectral response of the camera is found to be the highest for this particular wavelength. A notch filter provided beneath the dichroic filter protects the camera by completely blocking the reflected or scattered laser light. The sample using the objective is imaged over the camera's sensor with a tube lens. The high-speed camera operating at 264,000 frames per second (fps) is used to record the events that occur at microsecond timescales.

\bigskip

\begin{figure}[ht]
    \centering
    \includegraphics[scale = 0.6]{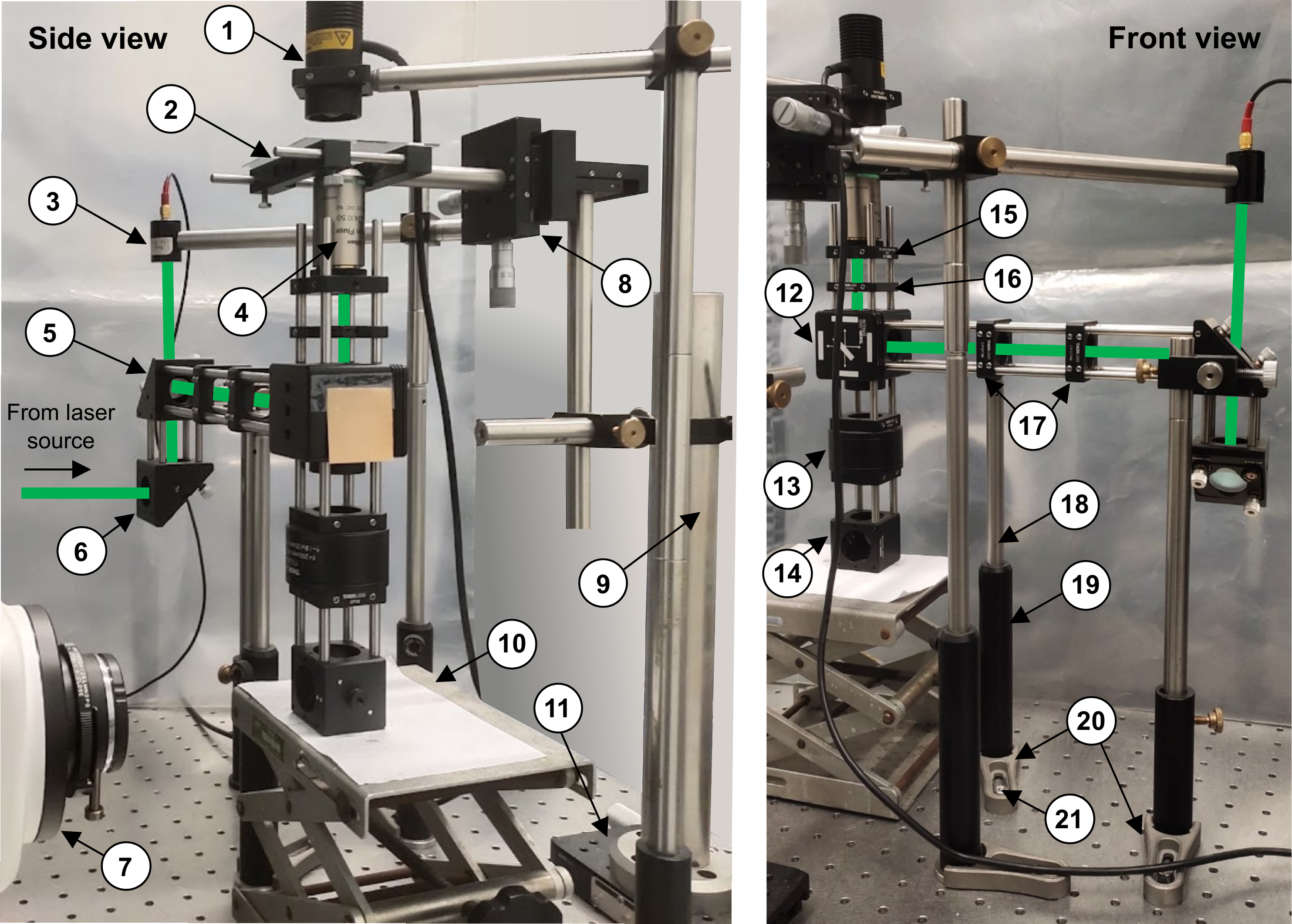}
    \caption{Annotation of components in the experimental setup. 1. Mounted LED, 2. Slide holder, 3. Energy sensor, 4. 20x objective, 5. Right-angle kinematic mirror mount with round beamsplitter, 6. Right-angle kinematic mirror mount with laser mirror, 7. Camera, 8. 3-axis travel stage, 9. Damped post, 10. Lab jack, 11. Linear translation stage 12. Kinematic fluorescence filter cube with dichroic and notch filters, 13. Tube lens with cage plates (tube lens) and threading adapters, 14. Cage cube with rectangular mirror, 15. Cage plate (objective) with threading adapter, 16. Cage system iris diaphragm, 17. Cage plates (lens), 18. Optical post, 19. Pedestal post holder, 20. Clamping fork, 21. Cap screw. The green track indicates the laser path.}
    \captionsetup{justification=centering}
    \label{fig: Physical setup}
\end{figure}

\noindent The laboratory view of the experimental setup is depicted in \Cref{fig: Physical setup}. The 3-axis translation stage (13 mm range with a least count of 10\,$\mu$m) that holds the sample is mounted over a linear translation stage (50 mm range and with a least count of 10\,$\mu$m) to have a larger travel range along the microchannel axis. The microchannels are fabricated using PDMS and offer several advantages over commercially available plastic microfluidic chips \cite{10.1002/elps.201100482}. For instance, the permeability of PDMS to gases can be exploited to remove residual gas bubbles and improve the functionality of the device in view of the cavitation experiments \cite{10.1039/C5LC00982K}. Furthermore, the elastomeric nature of PDMS can mimic vascular pressure-driven flows which can also aid in providing a real-time flow representation \cite{10.1109/TBCAS.2019.2946519}.

\bigskip

\noindent With the existing system, laser-induced cavitation is implemented in the microchannels using a solution of aqueous red dye (DR81, Sigma-Aldrich). The addition of a doublet lens in the laser path before the dichroic filter enables us to perform fluorescence-PIV as a second application. PIV is carried out in a pressure-driven flow using fluorescent tracer particles (0.955 $\mu$m, Rhodamine B). The cavitation and PIV experiments performed in this work are independent of each other and only highlight the adaptability in the design of the optical setup. However, these can also be combined together into a single experiment depending on the nature of the study \cite{10.1038/srep05445,10.1039/B806912C}.


\bigskip

\bigskip

\section{Design files summary}

The design files summary contains the required CAD and STL files that are necessary for 3D printing the mold of microchannels.

\begin{table}[ht]
  \small
  \centering
    \begin{tabular}{|>{\centering\arraybackslash}p{16em}|>{\centering\arraybackslash}p{12em}|>{\centering\arraybackslash}p{10em}|>{\centering\arraybackslash}p{10em}|} \hline
    \textbf{Design filename} & \textbf{File type} & \textbf{Open source license} & \textbf{Location of the file} \\ \hline
    Microchannels mold (for 3D printer) & SLDPRT, STEP and STL & CC BY 4.0 & \mendeley \\
    \hline
    \end{tabular}%
\end{table}%

\section{Bill of materials summary}



The spreadsheet containing the bill of materials (\textbf{Bill of materials.xlsx}) can be found in the repository: \href{http://dx.doi.org/10.17632/fy86gxm4mp.1}{\nolinkurl{http://dx.doi.org/10.17632/fy86gxm4mp.1}}. The materials are mentioned in the same order as presented in the following section. An additional spreadsheet containing the description of the materials (\textbf{Description of materials.xlsx}) used is also provided in the repository. Please note that the material costs furnished might vary with different suppliers, time period and the location of purchase.

\section{Build instructions} \label{Build instructions}

This section provides the complete procedure for the assembly of the experimental setup using discrete optomechanical components. The main emphasis is laid on the ease of alignment and the quality of imaging. The optical train is constructed predominantly using 30 mm cage components. The details of mountings for each of the components are given in the respective links provided in the bill of materials. The setup is built starting with the dichroic filter and subsequently branched in three directions namely the objective branch, tube lens branch and the laser guidance and telescope branch as depicted in \Cref{fig: microscope construction}. Additionally, some laser mirrors are used to guide the laser path from the source and this might vary in each case.

\subsection{Microscope construction}
\label{Microscope construction}

\begin{figure}[ht]
    \centering
    \includegraphics[scale = 0.9]{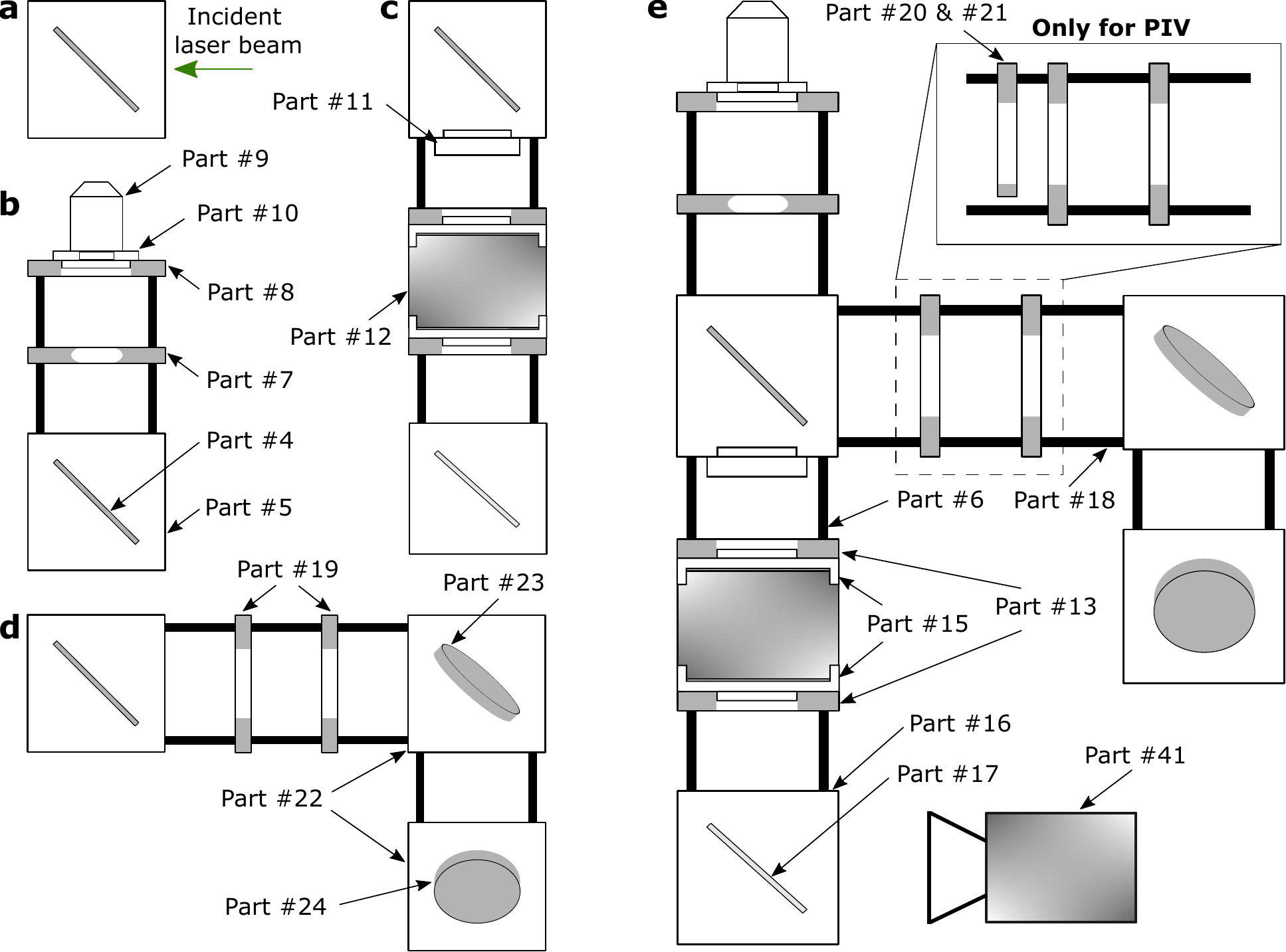}
    \caption{Sketches of a) dichroic filter in a cage system. b) sub-assembly of the objective branch. c) sub-assembly of the tube lens branch. d) sub-assembly of laser guidance and telescope branch. e) complete assembly of the microscope.}
    \captionsetup{justification=centering}
    \label{fig: microscope construction}
\end{figure}

\begin{enumerate} [listparindent=\parindent]
    \item At first, mount the laser (Part \#1) on the optical breadboard (Part \#2) with the help of Allen keys (Part \#3) while allocating sufficient space for the remaining components of the setup.
    
    \item Place the dichroic filter (Part \#4) inside the kinematic fluorescence filter cube (Part \#5) and position it accordingly such that it is inclined at 45$^\circ$ to the incident laser beam while the coated side faces the incoming laser light as shown in \Cref{fig: microscope construction}a.

    \item Using the help of cage assembly rods (Part \#6), assemble the cage system iris diaphragm (Part \#7) followed by the cage plate (Part \#8) for the objective on the side of the reflected laser light as depicted in \Cref{fig: microscope construction}b.
    
    \item Screw the 20x objective (Part \#9) to its cage plate while making use of the threading adapter (Part \#10) in order to match the screw threads. In the current setup, the objective has a NA = 0.5 and focal length, $f_\mathrm{objective}$ = 10 mm. The calculated pupil diameter of the objective is, $2\, (\tu{NA})\,f_\mathrm{objective} = 10 \,\tu{mm}$. 

    \item On the opposite side of the kinematic fluorescence filter cube, secure the notch filter (Part \#11) in the emission filter port with the engraved transmission arrow pointing away from the dichroic filter.

    \item Following this, sandwich the tube lens (Part \#12) between the two cage plates (Part \#13) with the aid of threading adapters (Part \#15) and then attach this sub-assembly to the cage assembly rods (as indicated in \Cref{fig: microscope construction}c) beneath the notch filter making sure that the transmission arrow points away from the notch filter. The required magnification ($m$) can be achieved by choosing the tube lens with the right focal length ($f_\mathrm{tubelens}$),
    \begin{equation}
        m = f_\mathrm{tubelens} / f_\mathrm{objective}.
    \end{equation}

\noindent In order to achieve a magnification of 20x, the tube lens with $f_\mathrm{tubelens} = 200\, \tu{mm}$ is chosen.

    \item At the bottom, connect the cage cube (Part \#16) containing the rectangular mirror (Part \#17) inclined at 45$^\circ$ to the incident light with the cage assembly rods.

    \item On the laser incident side of the kinematic filter cube, fasten the cage assembly rods (Part \#18) and attach two cage plates (Part \#19) for lenses as in \Cref{fig: microscope construction}d. The use of lens is required only if there is a need for varying the laser source beam diameter as shown in \Cref{fig: Expt setup}. One of the cage plates is secured whereas the other is left free to slide so that adjustments can be made depending on the focal length of the lenses used. Additionally, a drop-in cage mount (Part \#20) holding the achromatic doublet lens (Part \#21) is mounted as shown in \Cref{fig: microscope construction}e only while performing the PIV experiments.

    \item Take a couple of right-angle kinematic mirror mounts (Part \#22) each encompassing a round beamsplitter (Part \#23) and a round laser mirror (Part \#24) respectively and connect them using the cage assembly rods. Insert this sub-assembly to the other end of the cage assembly rods extending from the kinematic filter cube as represented in \Cref{fig: microscope construction}d. The distance between the two mirror mounts can be adjusted for varying the height based on the laser source position by adding or removing the cage assembly rods.

    \item In order to erect the setup upright, secure the right angle kinematic mirror mount to the optical breadboard with the help of optical posts (Part \#26), right angle post clamp (Part \#27), pedestal post holder (Part \#28), clamping fork (Part \#29) and cap screw (Part \#30). Place the lab jack (Part \#31) beneath the cage cube for support as portrayed in \Cref{fig: Physical setup}.

    \item Alongside the setup, clamp the linear translation stage (Part \#32) to the breadboard and mount the damped post (Part \#33) on top of it (refer \Cref{fig: Physical setup} - side view). Insert the post mounting clamp (Part \#34) through the damped post and tighten it. By making use of the optical posts and right angle post clamp, set up the 3-axis travel stage (Part \#35). Connect the slide holder (Part \#36) to the travel stage using an optical post and ensure that it remains on top of the objective by adjusting the height. The linear travel stage facilitates a larger translation in the horizontal direction whereas the 3-axis travel stage helps with smaller translations in all three directions. In addition, the post mounting clamp provides a rotational degree of freedom to the slide holder.

    \item With the support of an optical post, post holder, clamping fork and cap screw, place the mounted LED (Part \#37) on top of the slide holder (refer \Cref{fig: Physical setup} - side view). Connect the cable from the LED light to the LED driver (Part \#38) and power the LED driver using the power source. The LED driver controls the light intensity.

    \item Similarly, align the energy sensor (Part \#39) above the beamsplitter to measure the energy of each laser pulse. The output of the sensor is connected to the oscilloscope (Part \#40) which reads the measured output voltage. The energy can be calculated from the voltage using the calibration curve of the energy sensor provided by the manufacturer for the laser wavelength used.

    \item Finally, position the camera (Part \#41) in-line with the cage cube as shown in \Cref{fig: microscope construction}e. Link the camera and laser by means of a BNC cable (Part \#42) to the digital delay generator (Part \#43). The camera is triggered to record starting with 5 image frames before pulsing the laser so as to image the initial condition of the sample just before laser irradiation.
\end{enumerate}

\subsection{Alignment of the optical components}
\label{Alignment of the optical components}

During the construction of the microscope, the objective, tube lens and camera are initially positioned based on the design specifications provided by the manufacturer. The working distance of the tube lens is used as the distance between the tube lens and the camera. While, the distance between the objective and tube lens, $L$, is calculated using \cite{pupildistance},
\begin{equation}
    L = f_\mathrm{tubelens}(\phi_2-\phi_1) / \phi,
\end{equation}
\noindent where $\phi_1$ is the objective exit pupil diameter, $\phi_2$ is the tube lens entrance pupil diameter and $\phi$ is the camera's sensor size.\\ 

\begin{figure}[ht]
    \centering
    \includegraphics[scale = 0.9]{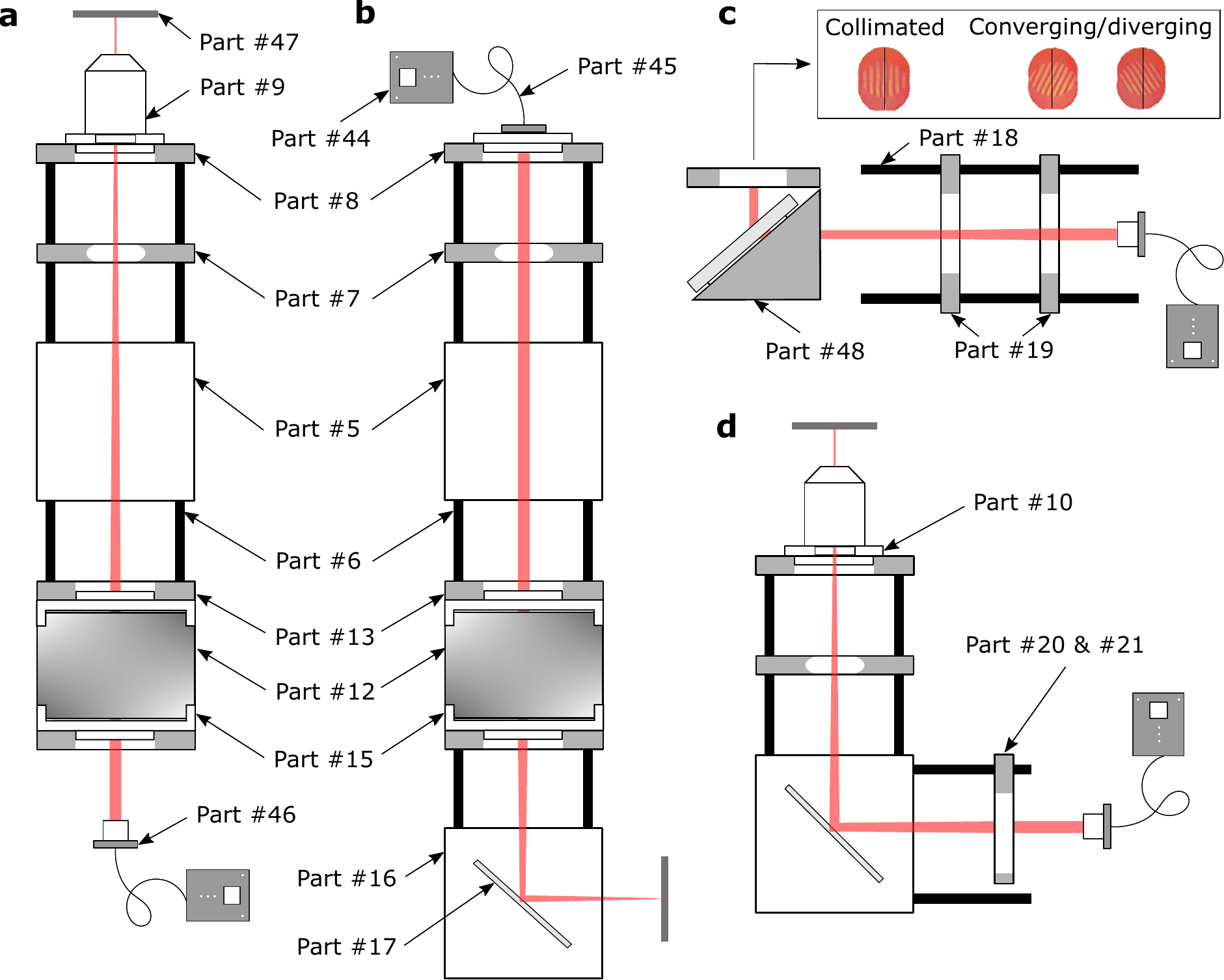}
    \caption{Illustrations of a) the alignment of objective and tube lens. b) the alignment of camera. c) the alignment of lenses. d) the alignment of doublet lens for PIV.}
    \captionsetup{justification=centering}
    \label{fig: microscope alignment}
\end{figure}

\noindent However, in practice, an alignment laser (Part \#44) is used for the purpose of aligning these components precisely \cite{10.1002/9783527671595}. This laser is first passed through the tube lens towards the objective as shown in \Cref{fig: microscope alignment}a. During this process, the dichroic and notch filters are removed temporarily so as to allow the transmission of the laser. The position of either the objective or the tube lens is varied such that the alignment laser exiting the objective's front lens remains collimated at any distance further from the objective. This is ensured with the help of a detector card (Part \#47). The above procedure allows us to determine the exact position of the tube lens relative to the objective or vice versa. The same alignment procedure can also be followed for non-infinity corrected objectives. However, a very precise positioning of the objective relative to the tube lens is required to overcome the drawbacks due to potential loss in image quality. On the other hand, infinity-corrected objectives allow the positioning of additional optical components such as dichroic and notch filters between the objective and camera without causing any aberrations or distortions.

\bigskip

\noindent In a similar manner, the precise position of the camera is determined to obtain a sharp image. First, the objective is unscrewed from the cage plate temporarily and the alignment laser is placed such that it passes through the kinematic fluorescence filter cube (without the dichroic filter) towards the direction of tube lens. The point where the tube lens focuses the alignment laser is found using the detector card as depicted in \Cref{fig: microscope alignment}b. The camera sensor is mounted exactly at this marked position. In this configuration, an additional mirror is used in between the tube lens and the camera to guide the light - allowing us to position the camera horizontally (stable orientation).

\bigskip

\noindent In the case of using a pair of lenses (as a Galilean telescope) to vary the source beam diameter, the lenses are initially placed at a distance equal to the sum of their focal lengths. In order to fix this distance accurately and ensure whether the beam remains collimated in the far field, the lenses are positioned in between the alignment laser and the shearing interferometer (Part \#48) as shown in \Cref{fig: microscope alignment}c. The laser light passing through the lenses forms an interference fringe pattern on the diffuser plate (or shear plate) of the shearing interferometer. When the fringes are parallel to the ruled reference line marked in the diffuser plate, the beam is collimated, and hence this distance between the lenses can be maintained while performing the experiments.

\bigskip

\subsection{Alignment for PIV experiment} \label{For PIV Experiments}
For performing the PIV experiment, a doublet lens is placed in between the right-angle kinematic mirror mount with the beamsplitter and the kinematic fluorescence filter cube containing the dichroic filter. The focal length of the doublet lens is chosen such that:

\begin{center}
    \textit{focal length of doublet lens = focal length of objective lens $\times$ demagnification,\\ demagnification = diameter of laser beam / field number},\\
\end{center}

\noindent where the desired \textit{field number} is the diameter of the view field. The position of the doublet lens is determined such that the beam from the alignment laser passing through the doublet lens remains collimated when it leaves the objective as represented in \Cref{fig: microscope alignment}d.

\subsection{Microfluidic device fabrication} \label{Microfluidic device fabrication}
Microfluidic devices allow us to both process and analyze samples with small volume handling. In this work, PDMS (Part \#49) is used to fabricate the microchannels \cite{10.1038/protex.2015.069}, for which a 3D printed mold is used. The microchannels with a rectangular cross-section ($100\,\mu$m\texttimes\,$250\,\mu$m) are employed. Since the dimensions of the channels described are in the order of hundreds of microns, a higher precision is required for preparing the mold. The mold is fabricated using Form 3+ resin printer (Part \#50) with a layer resolution of $25\,\mu$m using Formlab's clear V4 resin. Alternatively, the molds can also be prepared using silicon wafers employing the photolithographic technique. Although silicon wafers offer very high layer resolution ($\sim 2-5\,\mu$m), they are quite expensive ($\sim$\,€1000).



\bigskip

\begin{figure}[!t]
    \centering
    \includegraphics[scale = 0.9]{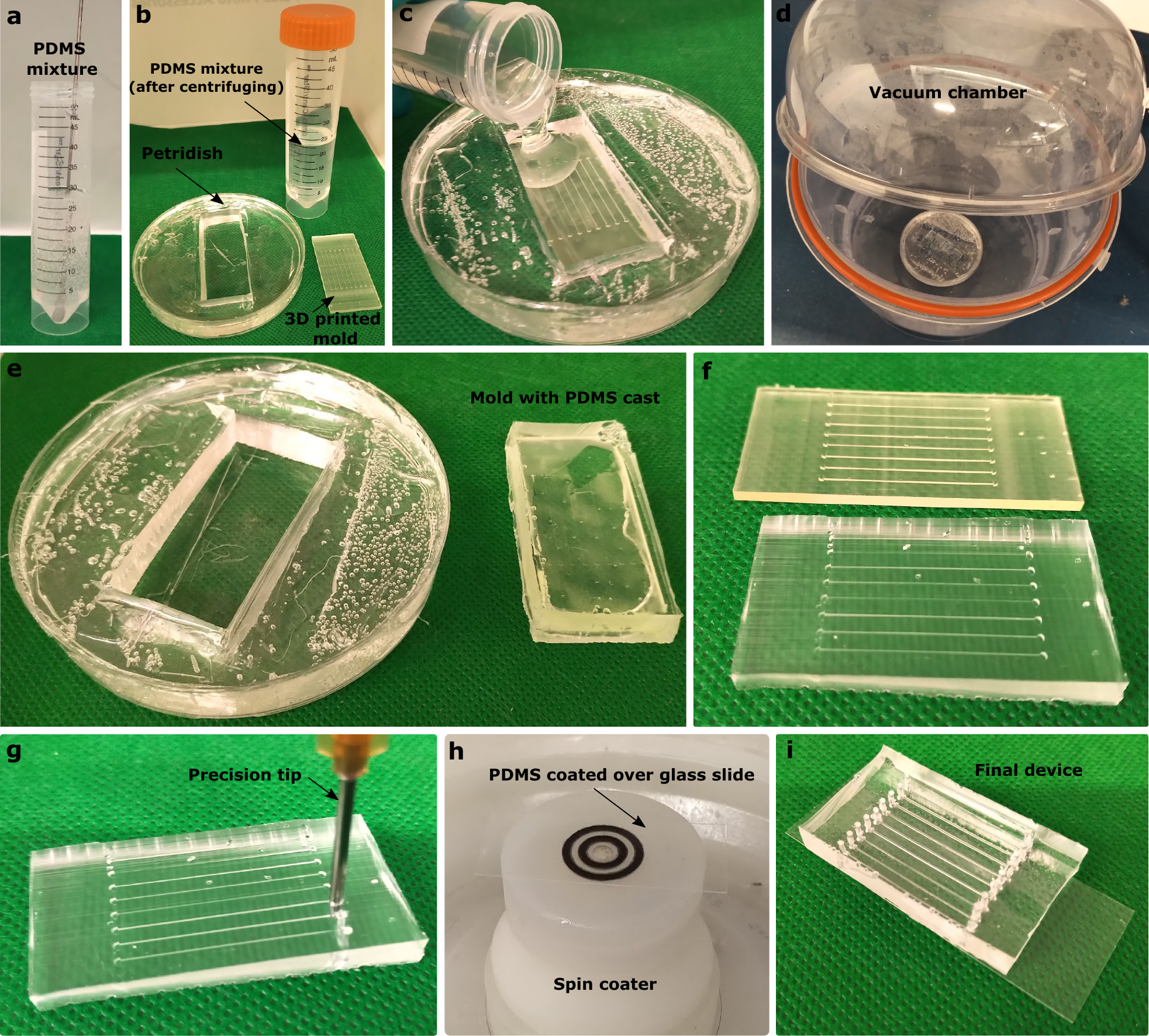}
    \caption{Visual demonstration of the fabrication procedure of microchannels. a) PDMS mixture is prepared. b) 3D printed mold is placed inside the petridish. c) The centrifuged PDMS mixture is poured into the petridish containing the mold. d) Vacuum is applied to remove the air bubbles present. e) PDMS is cured together with the mold and the required portion of PDMS along with the mold is cut carefully after curing. f) PDMS cast is separated from the mold. g) Inlet and outlet ports are created using the precision tip. h) A layer of PDMS is spin-coated over the cover glass. i) Coated cover glass is bonded with the PDMS cast to obtain the final device.}
    \captionsetup{justification=centering}
    \label{Microchannels Fabrication}
\end{figure}

\noindent The microchannels preparation process is illustrated in \Cref{Microchannels Fabrication}. During the fabrication, a cover glass is attached to the PDMS cast to cover the open side of the channels. For this purpose, the PDMS mixture is applied over a \#1 cover glass (Part \#51) using a spin coater (Part \#52), which is then bonded to the cast. The inlet and outlet ports for the microchannels are created by piercing the PDMS cast with a precision tip (Part \#53) before attaching the cover glass. A detailed preparation protocol can be found in the \textbf{Microchannel preparation.docx} document.\\

\noindent During the experiment operation, the laser passes through both the cover glass and a thin layer of PDMS within the microchannel. The PDMS coating over the cover glass in the microchannel preparation is not only used to bond the PDMS cast to the cover glass, but also ensures that the fluid experiences uniform friction across all the surfaces that it is in contact with.

\section{Operation instructions} \label{Operation instructions}
\underline{Safety}: Always remember to wear the laser safety glasses (Part \#54) to protect your eye during the operation of the laser.

\subsection{Laser-induced cavitation in microchannels} \label{Laser-induced cavitation in a microchannel}

Transient micro-vapor bubbles are produced by focusing the laser within aqueous solutions containing a red dye. Since water is transparent to 532 nm wavelength, a dye is used to facilitate the solution's absorbance to produce thermo-cavitation \cite{10.1039/C0LC00520G}. The red dye (Part \#55) with a concentration of 0.5 wt\% and an absorption coefficient of 173.47 cm$^{-1}$ is used in this study.

\bigskip

\noindent The laser (Nano L 50-50 PIV, Litron) is operated in a pulsed mode using the delay generator to produce single laser pulses on demand. The flash lamp and Q-switch are triggered externally using two TTL pulses (4V) with $96\,\mu$s time delay. This delay between the flash lamp and Q-switch can vary based on the laser model and the oscillator (for twin-headed lasers). The digital delay generator is set to externally trigger the camera at least 5 image frames before the Q-switch is triggered. Before pulsing the laser, the settings for the camera which include the field of view in pixels, frame rate and trigger options for recording are assigned accordingly using the Photron FASTCAM Viewer (PFV) 4 software (\href{https://photron.com/photron-support/}{\nolinkurl{https://photron.com/photron-support/}}). After the laser pulse is shot, the created vapor bubble exists for tens of microseconds depending on the energy supplied. The recorded images are saved in `.MRAW' file format as it can then be converted to any desired format using the same software. Further details regarding the absorbance measurements, calibration of the camera and experiment protocols for sample preparation and setup operation can be found in Raghunathan \cite{raghunathan2022exploring}.

\subsection{Visualization of the flow field in microchannels}

Fluid flow inside microchannels is visualized by performing fluorescence-PIV. Tracer particles (fluorescent indicators) are prepared with 2.5\% w/v aqueous dispersion of Rhodamine B (0.955\,$\mu$m with COOH=80 $\mu$mol\,g$^{-1}$, Part \#56) by diluting in water to 0.09 v/v\%. The closest alternative to the Rhodamine B particles (with a diameter of 0.955\,$\mu$m) currently available in the market is the red-fluorescent particles (Part \#57).\\

\noindent The operation procedure for the laser and camera similar to the cavitation experiment is adopted with a few modifications as follows. During these experiments, the laser is triggered externally with the triggering rate set to 50 Hz - the maximum allowable frequency for the laser. In coordination with the laser, the camera's recording frequency is also set to 50 fps. The recording time in the camera is set to 2\,s. Pressure is applied to one end of a channel to induce flow and is regulated using a flow control system (Part \#58). Real-time control and automation of pressure is achieved using the software interface OxyGEN (\href{https://www.fluigent.com/research/software-solutions/oxygen/}{\nolinkurl{https://www.fluigent.com/research/software-solutions/oxygen/}}). The top and bottom walls of the microchannel are determined by identifying the stationary particles.

\section{Validation and characterization} \label{Validation and characterization}

Following the instructions for constructing the microscope and fabricating the optofluidic device, we demonstrate the working and validation of the setup. Here, we exploit two different implementations of the hardware: laser-induced cavitation and fluorescence-PIV, achieved with minimal modifications to the microscope assembly. The validation of this working setup along with the customizations can prove to be beneficial for a multitude of research applications \cite{10.48550/arxiv.2301.09434}.


\subsection{Bubble dynamics}

\begin{figure}[htbp]
    \centering
    \includegraphics[width=\textwidth]{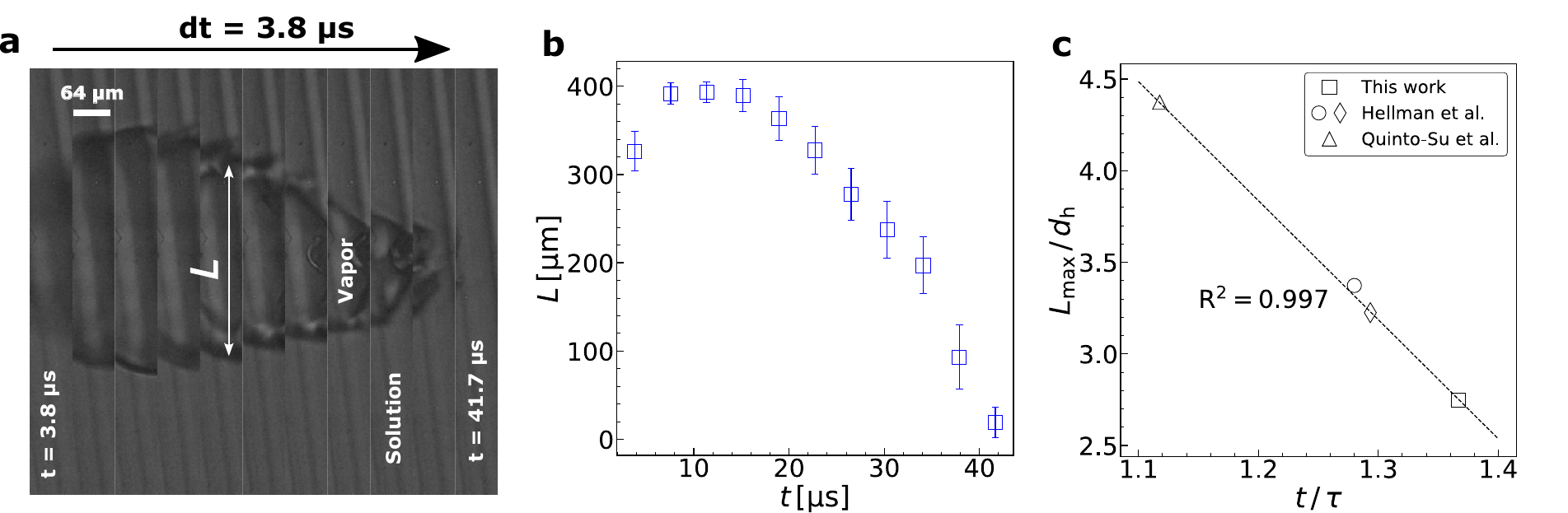}
    \caption{Laser-induced cavitation experiments. (a) Experimentally obtained bubble images recorded using the high-speed camera operated at 264,000 fps and supplied laser pulse energy of 118.1 $\mu$J. (b) Bubble evolution obtained from the post-processed experimental images. The error bars represent the standard error of 5 trials performed. (c) The non-dimensionalized parameter values compared against experiments from literature \cite{10.1021/ac070081i,10.1039/B715708H}. $L_\mathrm{max}$ represents the maximum size of the bubble.}
    \captionsetup{justification=centering}
    \label{Cavitation experiments}
\end{figure}

The microchannel filled with aqueous red dye solution is placed on the slide holder and the laser pulse at an energy of 118.1 $\mu$J is focussed at the channel centre to produce a vapor bubble. The images of the vapor bubble captured using the high-speed camera are represented in \Cref{Cavitation experiments}a. Only the central part of the bubble is imaged due to lower number of pixels available for recording at the operated high frame rate.\\ 

\noindent The time evolution of the vapor bubble after processing the images is presented in \Cref{Cavitation experiments}b. The size of the bubble ($L$) and the time ($t$) are non-dimensionalized with respect to the channel hydraulic diameter ($d_\mathrm{h}$) and collapse time of the bubble ($\tau$), defined as the time taken by the bubble to collapse from maximum size to zero, respectively. This non-dimensionalization allows us to compare our results with the existing literature. The observed time scale associated with the length scale of the bubble is in good agreement with the experiments reported by Hellman et al. \cite{10.1021/ac070081i} and Quinto-Su et al. \cite{10.1039/B715708H} as shown in \Cref{Cavitation experiments}c. The values of the parameters used to plot \Cref{Cavitation experiments}c are provided in \Cref{parameter values} below.
\begin{table}[htbp]
  \centering
  \caption{Parameter values used in the analysis of laser-induced cavitation experiments.}
    \begin{tabular}{>{\centering\arraybackslash}p{10em}>{\centering\arraybackslash}p{5em} >{\centering\arraybackslash}p{13em}>{\centering\arraybackslash}p{5em} >{\centering\arraybackslash}p{5em}} \hline
    \textbf{Author} & \textbf{Marker} & \textbf{Channel Dimensions} [$\mu$m] & $\mathbf{d_\mathrm{h}}$ [$\mu$m] & $\mathbf{\tau}$ [$\mu$s] \\ \hline
    This work & $\square$ & $100\times250$ & 143 & 30\\ 
    Hellman et al. \cite{10.1021/ac070081i} & $\bigcirc$ & $50\times100$ & 66.7 & 25\\ 
    Hellman et al. \cite{10.1021/ac070081i} & $\Diamond$ & $50\times200$ & 80 & 15\\ 
    Quinto-Su et al. \cite{10.1039/B715708H} & $\triangle$ & $30\times50$ & 37.5 & 17\\\hline 
    \end{tabular}%
    \label{parameter values}
\end{table}%

\subsection{Flow field detection}

\begin{figure}[htbp]
    \centering
    \includegraphics[width=0.9\textwidth]{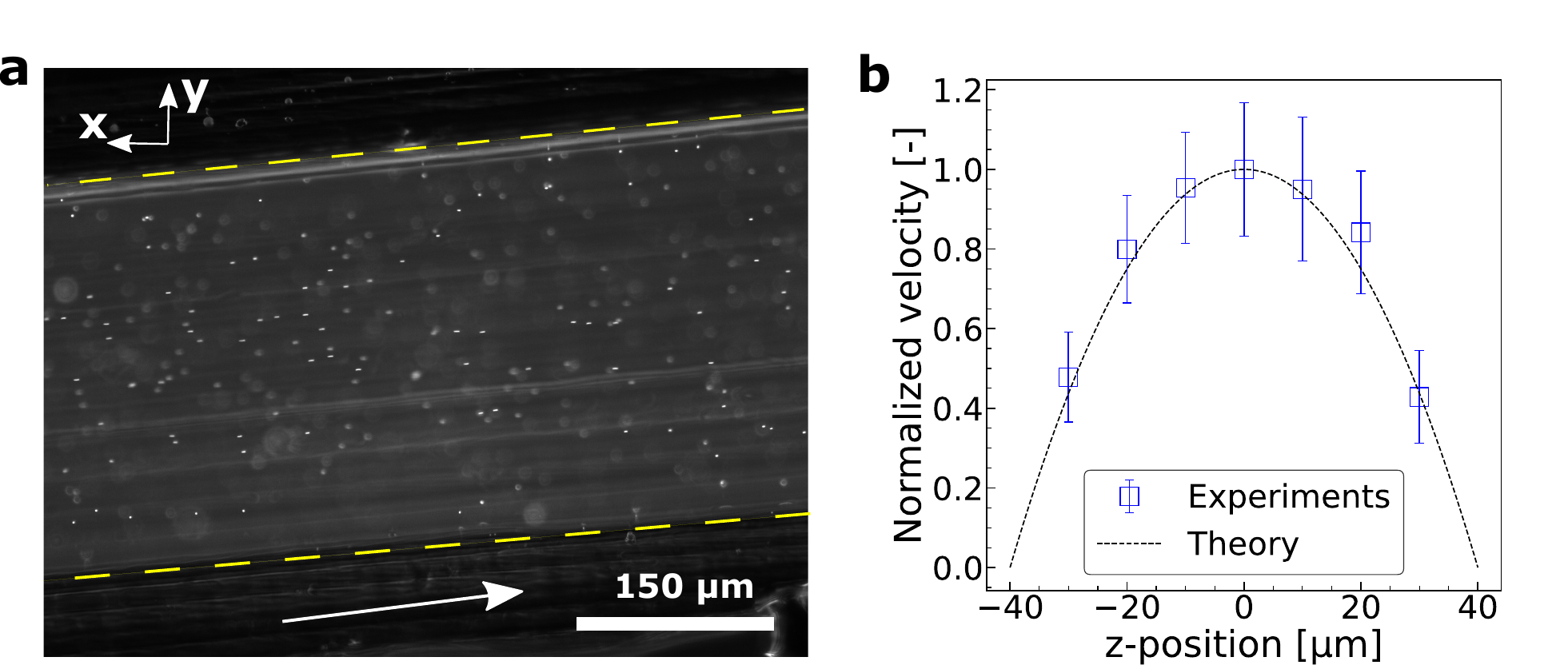}
    \caption{PIV measurements. (a) Typical example image from the PIV experiments performed. The white dots represent the fluorescent particles, the yellow lines indicate the channel boundaries (250 $\mu$m apart) and the white arrow points to the direction of the flow. (b) Normalized flow velocity along the x direction. $z=0$ is the mid-plane of the channel having a depth of 80 $\mu$m.}
    \captionsetup{justification=centering}
    \label{PIV measurement}
\end{figure}

A $100_{-25}^{+0}\,\mu$m \texttimes\, $250_{-30}^{+0}\,\mu$m microchannel is used to perform the PIV experiments. The flow of the fluorescent particles is imaged at different planes along the depth of the channel ($100\,\mu$m) starting from the channel bottom in steps of 10 $\mu$m. Since the frequency of the laser source is 50 Hz, the camera is also operated at 50 fps. An example image acquired for the flow field measurement through PIV within the microchannel is shown in \Cref{PIV measurement}a.

\bigskip

\noindent The data acquired from the PIV measurements is processed with the OpenPIV package (\href{http://www.openpiv.net/openpiv-python/}{\nolinkurl{http://www.openpiv.net/openpiv-python/}}) using the Python script \textbf{PIV analyzer.ipynb}. The velocity of the flow along the depth ($z$-direction) of the channel can be expressed using the following equation \cite{10.1039/C4SM00664J}:

\begin{equation}
    u = u_\mathrm{max} \left(1-\frac{4z^2}{H^2}\right)
    \label{Hele-Shae flow}
\end{equation}

\noindent where $H$ is the channel depth, $z \in [-H/2,H/2]$ is the distance from the mid-plane and $u_\mathrm{max}$ the magnitude of the velocity at $z=0$.

\bigskip

\noindent The normalized velocities measured at different distances from the microchannel's mid-plane are plotted in \Cref{PIV measurement}(b). It can be seen that the results obtained from the experiments are in very good agreement with the theory for shallow laminar flows (Hele-Shaw flows).\\

\noindent To conclude, we have provided a guide for building a low-cost microscope from scratch dedicated to laser-assisted optofluidic studies with fully customized optics. Two different experimental techniques were carried out using the working setup with minimal changes in the optical architecture. Furthermore, the in-house fabrication procedure for the microfluidic device that is used to analyze the sample solutions is also demonstrated. Finally, we have validated the experiments against the theory and empirical data to exhibit the full working functionality of the system. With dedicated adaptations, this proposed design can be utilized for a vast range of applications, e.g., in microfluidics \cite{doi:10.1063/5.0063714,doi:10.1021/acs.cgd.1c01436} and laser-induced crystallization research \cite{doi:10.1021/acs.cgd.0c01415}.\\

\noindent
\textbf{Ethics statements}\\
\indent \textbf{Animal and human rights}: Not applicable.\\

\noindent
\textbf{Credit author statement}\\
\indent \textbf{Nagaraj Nagalingam}: Conceptualization, Methodology, Validation, Investigation, Writing - Original Draft. \textbf{Aswin Raghunathan}: Methodology, Visualization, Writing - Original Draft. \textbf{Vikram Korede}: Methodology, Resources. \textbf{Edwin F. J. Overmars}: Resources. \textbf{Shih-Te Hung}: Methodology. \textbf{Remco Hartkamp}: Writing - Review \& Editing, Supervision. \textbf{Johan T. Padding}: Writing - Review \& Editing, Supervision. \textbf{Carlas S. Smith}: Methodology, Writing - Review \& Editing. \textbf{Huseyin Burak Eral}: Writing - Review \& Editing, Supervision, Funding acquisition.

\bigskip

\noindent
\textbf{Declaration of Competing Interest}\\
\indent The authors declare that they have no known competing financial interests or personal relationships that could have appeared to influence the work reported in this paper.

\bigskip

\noindent
\textbf{Acknowledgements}\\
\indent The authors thank Rumen Georgiev for the training in microfluidic device preparation. We also thank Dr. Daniel Irimia for guiding us with the optics. This work was funded by the Netherlands Science Foundation (NWO) through the Open Technology Programme, project number 16714 (LightX).

\printbibliography

\begin{minipage}{0.3\linewidth}
    \includegraphics[width=0.8\linewidth]{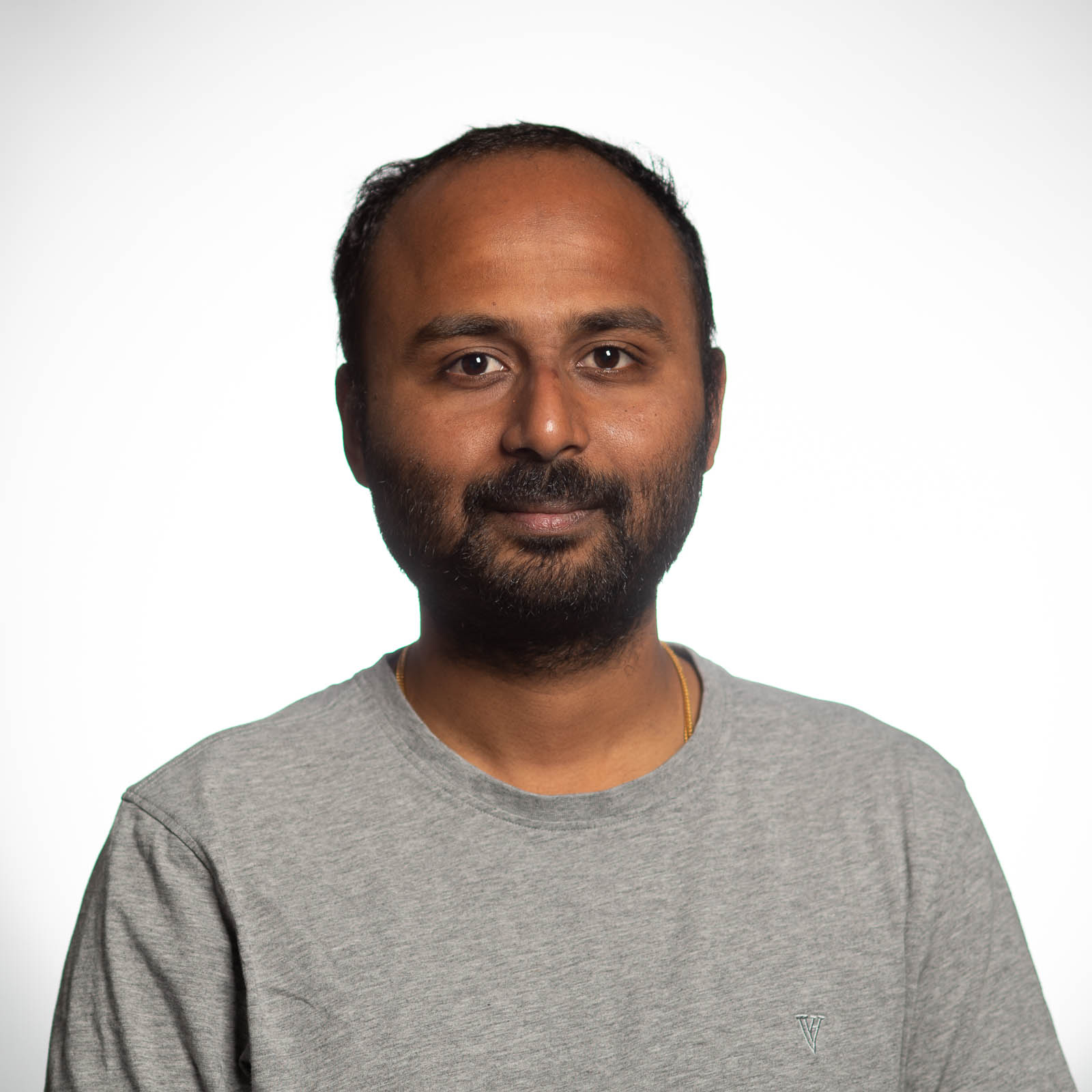}
\end{minipage}\hfil
\begin{minipage}{0.65\linewidth}
\small
\textbf{Ir. Nagaraj Nagalingam} is a PhD researcher in the Process \& Energy department at the Delft University of Technology, where he investigates the physico-chemical phenomenon of ``Non-Photochemical Laser Induced Nucleation'' through experiments and numerical simulations. Nagaraj received his M.Sc. degree in Mechanical Engineering – specializing in Energy, Flows and Process Technology from TU Delft. His previous work during the masters was on “Shape dependent motion of particles subjected to Stokes/creeping flow in a microfluidic channel having a Hele-Shaw geometry”. He is also a qualified Application Engineer in the field of turbochargers with 2 years of experience. He has a strong determination to contribute to the field of transport phenomena.
\end{minipage}\\\\

\begin{minipage}{0.3\linewidth}
    \includegraphics[width=0.8\linewidth]{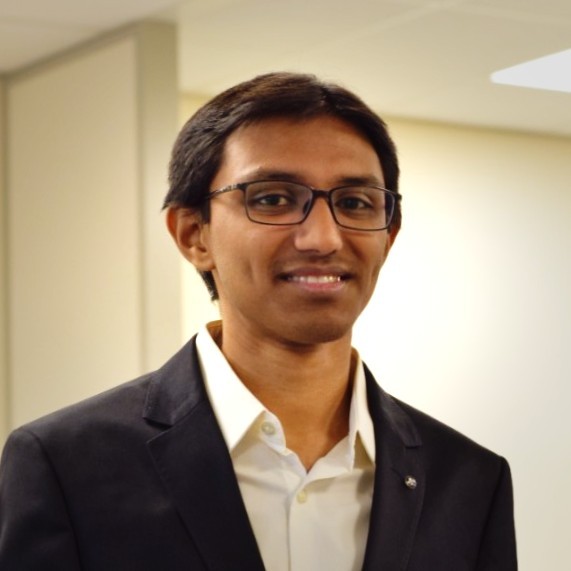}
\end{minipage}\hfil
\begin{minipage}{0.65\linewidth}
\small
\textbf{Ir. Aswin Raghunathan} is a master graduate from the Process \& Energy department at the Delft University of Technology, where he investigated the mechanism behind laser-induced crystallization through microfluidic experiments. Previously, he received his bachelors degree in Mechanical Engineering from SASTRA University, India in 2016. His research interest includes fluid mechanics applied to industrial processes. He also has over 2 years of work experience as a Design Engineer in the automotive sector.
\end{minipage}\\\\

\begin{minipage}{0.3\linewidth}
    \includegraphics[width=0.8\linewidth]{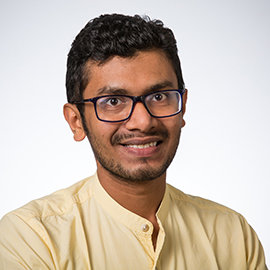}
\end{minipage}\hfil
\begin{minipage}{0.65\linewidth}
\small
\textbf{Ir. Vikram Korede} is a PhD researcher at the Process \& Energy department in the Delft University of Technology. He is working on non-photochemical laser induced nucleation and anti-solvent crystallization. Vikram received his M.Sc. degree from the Department of Chemical Engineering at TU Delft and bachelors degree from the National Institute of Technology, Tiruchirappalli, India. During his masters, he worked on the topic titled “Controlling crystal size in solvent exchange process”. He combines the knowledge of microfluidics, crystallization, optics, multiphase phenomena and aims to understand the interplay happening between light and matter in producing highly crystalline materials that has direct applications to fine speciality chemicals industries, pharmaceutical industries, catalyst industries, etc.
\end{minipage}\\\\

\begin{minipage}{0.3\linewidth}
    \includegraphics[width=0.8\linewidth]{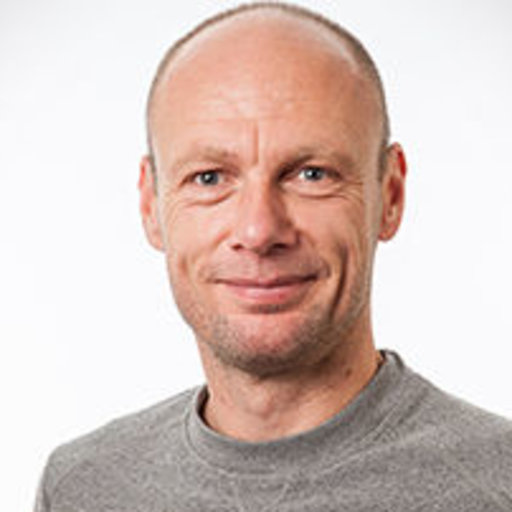}
\end{minipage}\hfil
\begin{minipage}{0.65\linewidth}
\small
\textbf{Ing. Edwin F. J. Overmars} is an expert in optical measurement techniques in the Laboratory for Aero and Hydrodynamics at Process \& Energy department at the Delft University of Technology.
\end{minipage}\\\\

\begin{minipage}{0.3\linewidth}
    \includegraphics[width=0.8\linewidth]{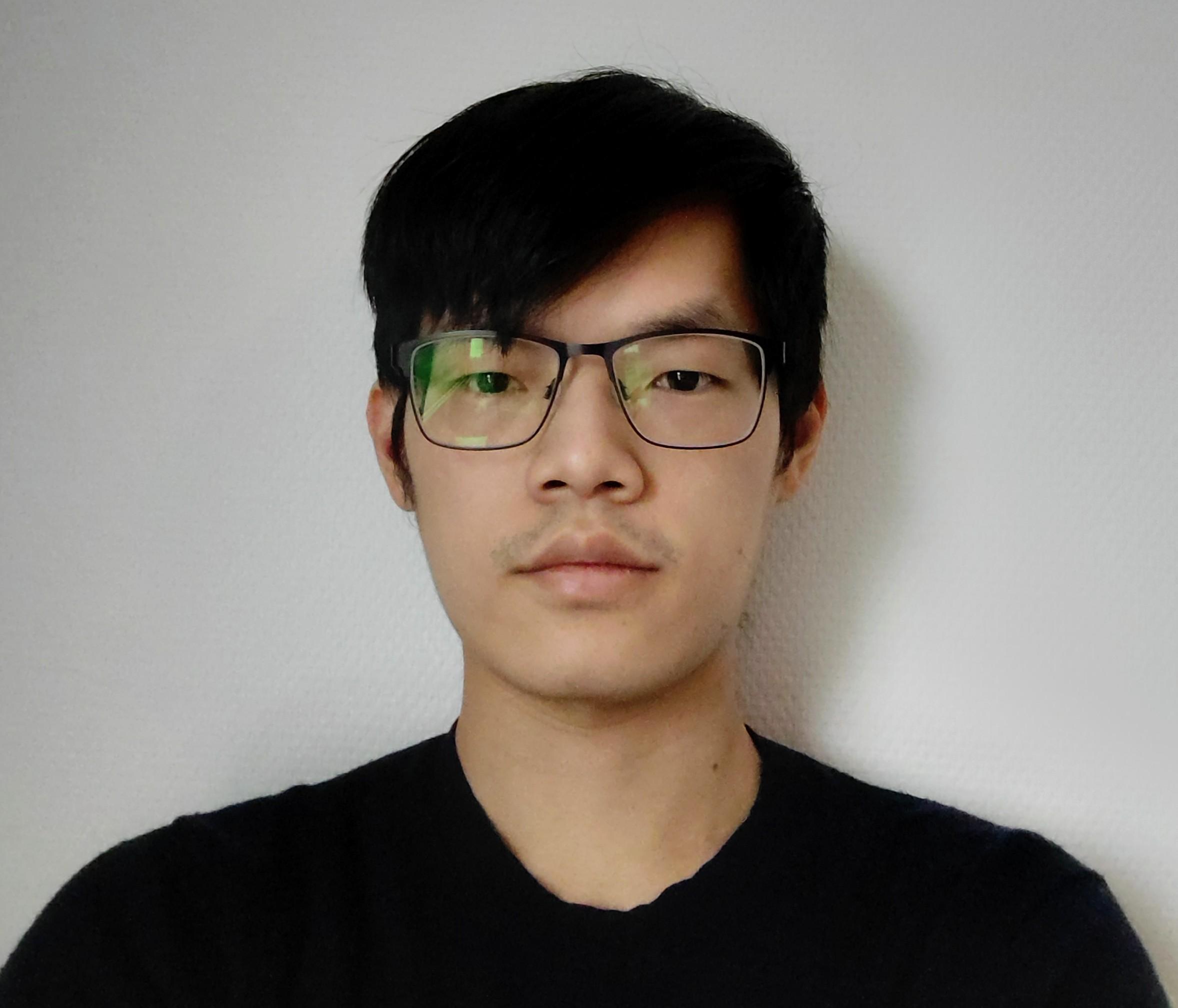}
\end{minipage}\hfil
\begin{minipage}{0.65\linewidth}
\small
\textbf{Shih-Te Hung} is a PhD student at Delft University of Technology. He received his Physics degree from NANU Taiwan, and a master’s degree in Photonics from Friedrich Schiller University, Jena, Germany. His research focuses on super-resolution microscopy and adaptive optics.  
\end{minipage}\\\\

\begin{minipage}{0.3\linewidth}
    \includegraphics[width=0.8\linewidth]{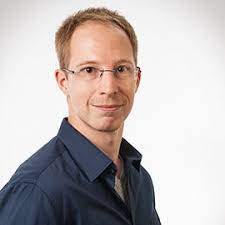}
\end{minipage}\hfil
\begin{minipage}{0.65\linewidth}
\small
\textbf{Dr. Remco Hartkamp} is an assistant professor in the Process \& Energy department at the Delft University of Technology, where he investigates molecular and macroscopic phenomena at the solid-fluid interface, with a special interest in the formation of electric double layers. Before starting his position at TU Delft in 2017, Dr. Hartkamp obtained a PhD (2013) in mechanical engineering from the University of Twente and Swinburne University of Technology and held postdoctoral positions at the Massachusetts Institute of Technology and Vanderbilt University. Website: \href{https://rmhartkamp.github.io/}{\nolinkurl{https://rmhartkamp.github.io/}}
\end{minipage}\\\\

\begin{minipage}{0.3\linewidth}
    \includegraphics[width=0.8\linewidth]{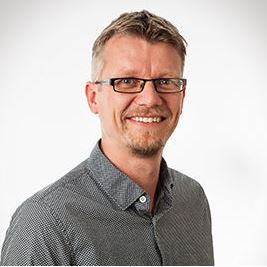}
\end{minipage}\hfil
\begin{minipage}{0.65\linewidth}
\small
\textbf{Dr. Johan T. Padding} is a full professor and chair of Complex Fluid Processing in the Department of Process and Energy (P\&E) at Delft University of Technology. He has an MSc in applied physics and a PhD in chemical physics from University of Twente, focusing on microscale and mesoscale simulations of soft matter. After several postdocs in Cambridge (UK), Twente (NL) and Louvain-la-Neuve (BE), Johan got a tenured position at Eindhoven University of Technology in 2011 where he shifted his attention to larger scale multiphase flows. He moved to Delft University of Technology in 2016. Website: \href{http://padding.freecluster.eu}{\nolinkurl{http://padding.freecluster.eu}}
\end{minipage}\\\\

\begin{minipage}{0.3\linewidth}
    \includegraphics[width=0.8\linewidth]{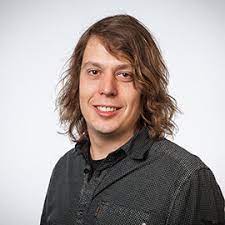}
\end{minipage}\hfil
\begin{minipage}{0.65\linewidth}
\small
\textbf{Dr. Carlas S. Smith} is an assistant professor in the Delft Center for Systems and Control at Delft University of Technology. His research transcends the classic boundaries between biology and engineering to visualize single molecules in living cells. For this purpose, he combines advanced control strategies, algorithm development, and computational microscopy technology.
\end{minipage}\\\\

\begin{minipage}{0.3\linewidth}
    \includegraphics[width=0.8\linewidth]{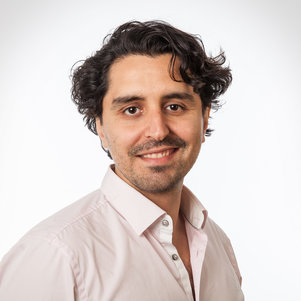}
\end{minipage}\hfil
\begin{minipage}{0.65\linewidth}
\small
\textbf{Dr. Huseyin Burak Eral} is an assistant professor in the Process \& Energy department at the Delft University of Technology, where he investigates crystal engineering, colloidal physics and hydrodynamics/microfluidics. After receiving his bachelor and master degrees in chemical engineering at Boğaziçi University and University of California Santa Barbara respectively, H. Burak Eral obtained his PhD degree in applied physics at the University of Twente in 2012. He went to the Massachusetts Institute of Technology to work as postdoc at MIT-Novartis center for continuous manufacturing. He then moved to Delft University of Technology in October 2016. He also holds a guest faculty position in Van’t Hoff labs in Utrecht University. Website: \href{https://erallab.com/}{\nolinkurl{https://erallab.com/}}
\end{minipage}

\begin{center}
---------------------------------------------------------------------------------------------------------------------------
\end{center}

\end{document}